# Phase matched second harmonic generation in an on-chip lithium niobate microresonator fabricated by femtosecond laser

Jintian Lin,[1] Yingxin Xu,[2] Jielei Ni,[3] Min Wang,[1,4] Zhiwei Fang,[1,5] Lingling Qiao,[1] Wei Fang,[2,*] and Ya Cheng[1,5,†]

[1]State Key Laboratory of High Field Laser Physics, Shanghai Institute of Optics and Fine Mechanics, Chinese Academy of Sciences, Shanghai 201800, China. [2]State Key Laboratory of Modern Optical Instrumentation, College of Optical Science and Engineering, Zhejiang University, Hangzhou 310027, China. [3]Department of Electrical and Electronic Engineering, The University of Hong Kong, Pokfulam Road, Hong Kong. [4]University of Chinese Academy of Sciences, Beijing 100049, China. [5]School of Physical Science and Technology, ShanghaiTech University, Shanghai 200031, China.

Correspondence and requests for materials should be addressed to W. F. (wfang08@zju.edu.cn); or Y. C. (ya.cheng@siom.ac.cn).

**Nonlinear optical processes in whispering gallery mode (WGM) microresonators have attracted much attention. Owing to the strong confinement of light in a small volume, a WGM microresonator can dramatically boost the strength of light field, thereby promoting the nonlinear interaction between the light and the resonator material. However, realization of efficient nonlinear parametric process in microresonators is a challenging issue. The major difficulty is to simultaneously ensure the phase matching condition and a coherent multiple resonance condition for all the waves participating in the nonlinear conversion process. Here, we demonstrate highly efficient second harmonic generation (SHG) in an on-chip lithium niobate microresonator fabricated by femtosecond laser direct writing. We overcome the difficulty of double resonance for the phase matched pump and second harmonic waves by selectively exciting high order modes in the fabricated thin-disk microresonator. Our technique opens opportunities for integrated classical and quantum photonic applications.**

## Introduction

Today, the widely adopted approaches to fabricate high quality (high-Q) WGM microresonators on semiconductor or dielectric chips are based on defining circular pads using lithographic method followed by formation of the microresonators via dry or wet etching[1,2,3]. In combination with $CO_2$ laser surface reflow, ultrahigh Q-factors on the order of $\sim 10^8$ have been demonstrated in silica microtoroid resonators which have benefited applications ranging from nonlinear optics and high-sensitivity biosensing to cavity quantum electrodynamics and optomechanics[3-5]. The purpose of $CO_2$ surface reflow is to smooth the surface of the microtoroid resonator fabricated in silica glass. The ultra-high surface smoothness resulted from surface tension of molten glass is the key ingredient for the high Q factors[3]. In comparison with glass materials, crystals can have higher nonlinear coefficients, broader transmission window, and higher damage threshold under exposure of intense light fields[6], making the crystalline microresonators particularly attractive for nonlinear optics and quantum light source applications[7-9]. Since most dielectric crystalline materials cannot be effectively processed with dry or wet etching, high-Q crystalline WGM resonators are typically fabricated using mechanical polishing[6]. Nevertheless, the mechanical polishing is not compatible with chip-level integration of microresonators into miniaturized photonic circuits and often limits the size of resonators to millimeter scale. It is therefore highly in demand to develop microfabrication techniques for producing on-chip dielectric crystal microresonators possessing both high-Q factors and small sizes[10-17].

Recently, we have demonstrated fabrication of high-Q lithium niobate (LN or $LiNbO_3$) microresonators on a LN thin film wafer[12,13]. The wafer is formed by bonding an ion-sliced LN thin film onto a LN substrate with a sandwiched silica layer[18]. The freestanding LN microdisk suspended on a silica pedestal is fabricated by femtosecond laser direct writing followed by focused ion beam (FIB) milling and selective chemical wet etching. The LN thin film has soon been used for mass-production of on-chip high-Q ($10^5$~$10^6$) LN microresonators[14,17]. Applications such as SHG and electro-optic modulation have been preliminarily demonstrated in the LN microresonators because of the large second-order nonlinear optical coefficient and electro-optic coefficient of LN crystal[10,12,14,17].

It is worth mentioning that in the nonlinear frequency conversion processes, phase matching often plays a determining role in terms of the achievable conversion efficiency[19,20]. Phase matching in bulk materials is typically realized using either critical phase matching (CPM) by arranging the angular crystal orientation with respect to the propagation direction of the incident light or quasi-phase matching (QPM) by patterning the optical nonlinearity of materials[20,21]. However, in general, it is not straightforward to achieve perfect CPM in a WGM microresonator because of the fact that when the waves circulate around the periphery of the microresonator, they can experience varying crystal orientation. The QPM is compatible with the nonlinear frequency conversion in WGM microresonators, which has indeed been demonstrated[22]. Nevertheless, to make use of the QPM mechanism, the additional nonlinearity patterning on crystalline microresonators leads to increased complexity particularly for microresonators of small sizes except for the crystals of $\bar{4}$-symmetry[22].

Recently, G. Lin et al. proposed a different phase matching scheme for SHG in a WGM resonator fabricated by mechanical polishing[23]. The scheme is coined as cyclic semi-phase matching (CSPM). In the CSPM, perfect phase matching takes place cyclically at four azimuth angles where the ideal angular configurations of the crystal and the propagating beams exist. The implementation of CSPM in a large WGM resonator is relatively easy for its high mode density, thereby facilitating both the fundamental and second harmonic waves to be tuned on resonance simultaneously. This has indeed been demonstrated by G. Lin et al in a beta barium borate resonator with a diameter of ~1.82 mm and a thickness of a few hundred microns[23]. Here, we demonstrate that the CSPM can also be employed for phase matched SHG in an on-chip LN microresonator with a diameter of ~100 μm. It should be stressed that for high-Q microresonators of small sizes, maintaining the coherent double resonance condition in the phase matched SHG is extremely challenging because of the sparse frequency distributions of the high-Q modes for both the fundamental and second harmonic waves. We overcome the difficulty by controlling the geometrical dimensions of the LN microdisk and selectively exciting certain high-order mode of the fundamental wave in the microresonator. Thanks to the low optical absorption and high nonlinear optical coefficient of LN crystal, we achieve a normalized conversion efficiency of $1.106×10^{-3}$/mW, which, to the best of our knowledge, is a record high in the normalized conversion efficiency of SHG as compared with the latest results of SHG experiments demonstrated in various on-chip WGM microresonators[12,14,22,24,25]. This is mainly enabled by achieving the double resonance through the selective excitation of high-order modes of the fundamental and second harmonic waves and successful implementation of the CSPM scheme in our high-Q on-chip LN microresonator with a diameter of only ~100 μm.

## Results

**Phase matched SHG in LN microresonator**

Phase matched SHG was carried out in an X-cut LN microresonator using an experimental setup as schematically shown in Fig. 1. The LN microresonator fabricated on an X-cut LN thin film has a thickness of 709 nm and a diameter of ~102.3 μm, as shown by its scanning electron microscope (SEM) image in the right-hand-side inset of Fig. 1. The optical axis of the crystal lies in the plane of LN microdisk. The details on the fabrication of LN microresonator can be found in **Methods**.

The second harmonic was generated using a continuous-wave (CW) tunable diode laser with a linewidth of 10 MHz (New Focus, Model: 688-LN) amplified by an erbium-ytterbium doped fiber amplifier (EYDFA) as the pump source. A fiber taper placed in contact with the top surface (*YZ* plane) of the microdisk was used for coupling the pump light into the LN microresonator. The same fiber taper was also used for coupling the second harmonic out of the LN microresonator. Selective excitation of a certain spatial mode can be achieved by adjusting the position of the fiber taper with respect to the microresonator. By using an online fiber polarization controller, the polarization of the pump laser can be chosen to enable excitation of TM (transverse magnetic) mode in the microdisk, i.e., the polarization of the pump light in the microresonator is perpendicular to the disk plane. The signals from the output port of the fiber taper were collimated by an objective lens with a numerical aperture (NA) of 0.25, and separated by a dichroic mirror. The generated second harmonic signal was measured with a photo-detector (PD, PD 1 in Fig. 1). Meanwhile, the pump light deflected by the dichroic mirror was measured with another PD (PD 2 in Fig. 1). More experimental details can be found in **Methods**.

Figure 2 (a) shows the measured spectrum of the second harmonic signal at a wavelength of 770.065 nm, which was obtained with a pump light whose spectrum is presented in Fig. 2(b) for comparison. The wavelength of the pump light is 1540.13 nm, which is exactly twice of that of the second harmonic signal. We notice that under this condition, the second harmonic beam is TE (transverse Electric) polarized whereas the fundamental beam is TM polarized. The procedures for determining the polarizations of both beams are described in **Methods**. As shown by the optical micrograph captured by a charge coupled device (CCD) camera in Fig. 2 (c), the scattering light of the second harmonic is clearly visible on the edge of the microdisk, which is the nature of WGM propagation.

We further measured the conversion efficiency of SHG as a function of the input laser power. It was observed that the conversion efficiency increases linearly with the increasing input laser power. From the slope of the fitting line in Fig. 2(d), a normalized conversion efficiency of $1.106\times10^{-3}$/mW can be determined. The result indicates that the combination of the high-Q LN microresonator and the CSPM mechanism can provide promising potential for highly efficient nonlinear wavelength conversion[23].

It is instructive to clarify whether the fundamental and second harmonic waves both satisfied the resonance condition in our experiment. For this purpose, we measured the transmission spectra around the wavelengths of fundamental TM and second harmonic TE waves coupled into the microresonator, as shown in Fig. 3(a) and (b), respectively. More details can be found in **Methods**. Indeed, from the spectra, we can confirm that both the fundamental and second harmonic wavelengths are on resonance with the modes in the LN microdisk. One can also evaluate the Q factors based on the spectra. At the pump wavelength around 1540.13 nm, the measured Q factor reaches $1.13\times10^5$ (see, inset of Fig. 3(a)). Around the spectral range of the second harmonic, the measured Q factor is ~$1.6\times10^4$ (see, inset of Fig. 3(b)).

## Discussion

The CSPM concept has been well explained in Ref. [23]. In brief, when the pump and second harmonic waves are TM and TE polarized respectively in the LN microdisk, and the optical axis ($Z$ axis) of the LN crystal is oriented in parallel with the disk plane, the pump wave will experience a fixed refractive index, whereas the refractive index is oscillating during the cyclical propagation of the second harmonic wave in the disk. Specifically, the effective refractive indices of both the pump and second harmonic waves can be engineered in the thin film microresonator with the thickness on the wavelength scale by tuning the geometrical dimensions and selectively exciting the spatial modes. The refractive index engineering is critical particularly for LN microresonator because of the large refractive index difference of the ordinary- and extraordinary-lights in the LN crystal. As we will show by numerical simulation below, after the above-mentioned optimizations, the fixed refractive index of the pump light and the oscillating refractive index of the second harmonic will cross at four azimuth angles cyclically in the microresonator. Perfect phase matching can be achieved near the crossing points.

To simulate the modes of fundamental and second harmonic waves in the microresonator as shown in the inset of Fig. 1, a finite element method is used. Details of simulation can be found in **Methods**. All the geometrical parameters of the LN microdisk were chosen as that measured from the fabricated microdisk. By comparing our simulation results with the measured transmission spectrum in Fig. 3(a), the fundamental mode is identified to be a high order mode of $TM_{6,359,1}$, where the subscript numbers (6,359,1) denote the radial mode number, azimuthal mode number, and axial mode number, respectively. The second harmonic mode is determined to be a high order mode of $TE_{1,712,3}$, as shown in Fig. 3(b). Due to the different mode numbers, the spatial overlap between the fundamental and second harmonic modes is calculated to be ~4.5% using an equation in Ref. [26]. Although the spatial overlap appears not very high, the reduction in the conversion efficiency caused by the insufficient overlap can be compensated by the high-Q factor of the microresonator that leads to extension of the interaction length.

Figure 4 (a) shows the calculated effective refractive indices of the fundamental and second harmonic modes as functions of azimuth angle $\theta$. These two curves cross at the four azimuth angles of 0.591, 2.550, 3.733, and 5.692, where the phase mismatch between the fundamental and second harmonic waves vanishes and the perfect phase matching condition is fulfilled, as depicted in Fig. 4(b). Similar to the effective refractive index of the second harmonic wave, the nonlinear coefficient in the X-cut LN microresonator also varies cyclically with azimuth angle, which is another determining factor in the SHG process. The calculated $\theta$-dependent nonlinear coefficient is presented in Fig. 4(c). It is found that among the four azimuth angles where the perfect phase matching is satisfied, the absolute values of nonlinear coefficient at the angles $\theta = 0.591$ and $\theta = 3.733$ are much higher than those at the other two azimuth angles. Therefore, we expect that within one cycle, efficient growth of the second harmonic signal will occur twice.

Finally, we calculated the field amplitude of the second harmonic wave as it grows along the periphery of the microresonator. The calculation was carried out by coherently integrating the locally generated second harmonic signal over the propagation trajectory. The details of the calculation is provided in **Methods**. Figure 4(d) shows how the second harmonic field evolves in the microresonator as a function of the azimuth angle. Because of the periodic characteristic of the double resonant SHG process, the calculation is carried out in the range of $0 \sim 2\pi$ for $\theta$. It is observed that in most angular range of $\theta$ where the phase matching is unavailable, the second harmonic signal oscillates rapidly without significant gain. The oscillation is a clear indication on the periodic exchange of the energies between the fundamental and second harmonic waves as a result of the phase mismatching. However, near the two azimuth angles $\theta = 0.591$ and

$\theta = 3.733$ where the absolute value of nonlinear coefficient is large and the perfect phase matching is fulfilled, the second harmonic signal increases dramatically to facilitate the efficient SHG as observed in our experiment. This behavior provides the essential microscopic mechanism behind the result in Fig. 2(d). We expect that extension of the CSPM mechanism to sub-100 μm crystalline microresonators can be effective for most negative uniaxial crystals.

We have demonstrated highly efficient SHG in a high-Q LN microresonator fabricated by femtosecond laser direct writing followed by FIB polishing. We overcome the challenges in simultaneously achieving phase matching and coherent double resonance by precisely controlling the geometry of the microresonator and selectively exciting certain high-order spatial modes. The demonstrated normalized conversion efficiency, which has reached $1.106\times10^{-3}$/mW, is a record high among the SHG experiments based on high-Q on-chip optical microresonators reported by far. For comparison, a conversion efficiency as high as $1.09\times10^{-4}$ mW$^{-1}$ was reported in Ref. [14], in which, however, the orientation of the LN crystal is different from that of the microdisk used in our work. Since the phase velocities of the fundamental and second harmonic waves strongly depend on the orientation of the birefringent LN crystal, the phase matching conditions are completely different in the two experiments. We attribute the one order of magnitude higher conversion efficiency observed in our experiment to the improved phase matching enabled by the CSPM scheme.

Our results provide further evidence on the applicability and effectiveness of the CSPM scheme for efficient wavelength conversion in crystalline microresonators, given that the coherent double resonance can be achieved in small-sized microresonators whose mode frequencies are sparsely spaced. The technique opens new possibilities for modern classical and quantum optical applications by enabling ultralow threshold, efficient on-chip nonlinear wavelength conversion. In principle, with the FIB fabrication, this technique provides sufficient resolution for fabricating an on-chip bus waveguide next to the microresonator. The will enable on-chip integration of a large number of microresonators with the bus waveguide. To this end, much effort is needed to optimize the key parameters, such as the gap distance between the waveguide and the microresonator, as well as the geometries of the microdisk and the waveguide to facilitate the selective excitation of the designated high-order WGM modes, which will be systematically investigated in the future.

## Methods

**Fabrication of X-cut LN microresonator**

In our experiment, commercially available ion-sliced X-cut LN thin film with a thicknesses of 0.709 μm (NANOLN, Jinan Jingzheng Electronics Co., Ltd) was chosen for fabricating the LN microdisk resonator. The LN thin film is bonded by a silica layer with a thickness of ~2 μm on an LN substrate[18]. Details in the process flow of fabrication of the LN microresonator have been described in Ref. [12,13], which can be summarized as the following 4 steps:

(1) The wafer was first immersed in water and ablated by femtosecond laser to form a cylindrical pad with a height of ~15 μm and a diameter of ~103 μm. Here, the femtosecond laser beam was focused into a tightly focused spot with a diameter of 1 μm for achieving high-precision ablation[27]. The average power of the laser beam was chosen as 0.35 mW when performing ablation in the LN material, and raised to 1 mW for ablation in the silica layer.

(2) The periphery of the cylindrical pad was polished by focused ion beam (FIB) milling to ensure a high Q factor of the fabricated microresonator. The diameter of the pad was reduced to ~102.3 μm after the process.

(3) A freestanding LN microdisk supported by a silica pedestal was then formed by selectively removing the silica layer under the LN thin film with diluted hydrofluoride acid (5%). The chemical wet etching lasted for 6 min.

(4) The sample was annealed at 500 °C for 4 hrs in air for further promoting the Q factor. We speculate that during the annealing process, the defects formed during the FIB polishing can be minimized in the thin film wafer, allowing for further reduction of the scattering loss.

**Second harmonic generation and characterization**

The experimental setup for investigating the SHG process in the fabricated LN microresonator is shown in Fig. 1. A narrow-linewidth tunable CW laser (New Focus, Model: 688-LN) was chosen as a seed of the pump source, whose tuning range is from 1510 to 1620 nm. The seed laser was amplified by the EYDFA to serve as the pump source, whose spectral range is from 1535 to 1570 nm. A fiber taper with a diameter of 700 nm pulled from a section of SMF-28 fiber was used to couple the pump light into the microresonator. An online fiber polarization controller was placed immediately before the fiber taper to control the polarization of the pump light. The relative position between the microresonator and fiber taper was controlled by an XYZ-piezo-stage. Two CCD cameras were arranged to simultaneously monitor both the side-view and top-view of the microresonators coupled with the fiber taper. To obtain the conversion efficiency of SHG, the power of the pump laser was first measured before it was coupled into the microresonator at the input port of the fiber taper; whereas the power of the second harmonic was measured from the output port of the fiber taper which was coupled to the microresonator after removing the fundamental light with a dichroic mirror followed by 3 pieces of shortpass filters. The transmission of the shortpass filter at ~1540 nm wavelength is 0.08%, whereas the transmission at ~770 nm wavelength is 99%.

We used a Glan-Taylor polarizer to examine the polarization states of fundamental and second harmonic waves in the microresonator in free space. A 10× objective lens with an NA of 0.25 was used to collect the scattering light from the microresonator. The spectra of scattered fundamental and second harmonic waves were measured by the spectrometer (Andor, Model Du920) after passing through a calibrated Glan-Taylor polarizer, as shown in the left-hand-side inset of Fig. 1. The width of slit mounted at the entrance port of the spectrometer was set to be 40 μm. An InGaAs detector was employed for detecting the scattered fundamental light at ~1540 nm. To detect the generated second harmonic wave at ~770 nm, a CCD array detector was used instead.

**Transmission spectra measurement**

To characterize the modes in the microresonator, the transmission spectra of the microresonator were measured using fiber taper coupling method[28] in the two wavelength ranges around 1540 nm and 770 nm, which correspond to the fundamental and second harmonic wavebands, respectively. The CW tunable laser diode (New Focus, Model: 688-LN) with a linewidth of 10 MHz and a scanning resolution of 0.1 pm was used to excite the modes around 1540 nm. Moreover, the same wavelength-tunable laser was first amplified and then frequency doubled to excite the modes around 770 nm. To avoid any nonlinear effects, the input powers of both the laser beams were set to be less than 0.01 mW. For the fundamental waveband, a transient optical power detector (model: 4650, dBm Optics Inc.) was used to measure the transmission spectrum from the output port of the fiber taper. The light signal in the spectral range from 1510 nm to 1620 nm can be recorded with 0.6 pm wavelength resolution and 0.015 dB power accuracy by the detector in less than 1 second. For the second harmonic waveband, a CCD array detector of the spectrometer was used to record the transmission spectrum.

To be consistent with our SHG experiment, TE-polarized modes in the microresonator were excited using the online fiber polarization controller for the second harmonic waveband, whereas TM-polarized modes were excited for the fundamental waveband.

**Simulations**

To identify the WGMs in the LN microresonator, simulations were carried out based on a finite element method (COMSOL Multiphysics, version 4.4) without inclusion of the fiber taper. For an axially symmetric microresonator, the frequencies and fields of the WGMs can be calculated by solving two-dimensional partial differential-equations[29]. The refractive indices used in the simulation were calculated using Sellmeier coefficients for congruently grown LN[30]. The ordinary indices are ~2.21 and ~2.26 at the fundamental and second harmonic wavelengths, respectively. The extraordinary index is ~2.18 at the second harmonic wavelength. For the TM modes, where the polarization is perpendicular to the optical axis of the crystal, the ordinary refractive index $n_o(\lambda)$ of bulk LN material was used. However, for the TE modes, the wave inevitably experiences the varying refractive index $n(\theta)$ along the periphery of the LN microresonator, which oscillates between the effective ordinary value $n_o(\lambda)$ and extraordinary value $n_e(\lambda)$ as below,

$$\frac{1}{n^2(\lambda,\theta)} = \frac{\cos^2\theta}{n_o^2(\lambda)} + \frac{\sin^2\theta}{n_e^2(\lambda)}. \tag{1}$$

In this case, an average refractive index can be obtained by integrating Eq. (1) with respect of the angle $\theta$ over 0 to $2\pi$, which is given below:

$$n_{av} = \left(\frac{1}{2}\left(\frac{1}{n_o^2(\lambda)} + \frac{1}{n_e^2(\lambda)}\right)\right)^{-1/2}. \tag{2}$$

For the thin film microresonator, the refractive indices calculated using the above equations cannot be directly used and should be further replaced by the effective refractive indices which depend on the geometry of the LN microresonator as well as mode spatial distributions. The effective refractive index $n_F$ of the TM-polarized fundamental mode was calculated using the propagation constants of the WGMs[31]. The TE-polarized second harmonic mode experienced a $\theta$-dependent oscillation between the effective ordinary value $n'_0$ and extraordinary value $n'_e$ during circulating along the periphery. The effective ordinary (extraordinary) refractive index around wavelength of 770.065 nm could be estimated from the TE-polarization WGMs in a microresonator which has a geometry exactly the same as the LN microresonator used in our SHG experiment, however with a constant ordinary (extraordinary) refractive index. We note that the average refractive index in Eq. (2) is only used for determining the mode number of the second harmonic field but not for calculating the difference in the wave vectors of the fundamental and the second harmonic wave (see, Eq. (4) below).

The second order nonlinear coefficient along the periphery in the SHG process can be calculated by the equation[32]:

$$d_{eff} = -d_{22} \cos\theta + d_{31} \sin\theta, \qquad (3)$$

where $d_{22}$ and $d_{31}$ are the second-order nonlinear coefficients for LN material. We assumed $d_{22} = 2.1\ pm/V$ and $d_{31} = -4.35\ pm/V$ in this calculation[33]. Clearly, the effective nonlinear coefficient varies cyclically as a function of the azimuth angle $\theta$. With undepleted pump approximation, growth of the second harmonic field $E_{SH}$ in the microresonator is described by the equation[32]:

$$\frac{dE_{SH}}{dz} = \frac{2i\omega_{SH}^2}{k_{SH}c^2} d_{eff} E_F^2 \cdot \exp\left(i \int \Delta k dz\right) - \frac{E_{SH}}{2k_{SH}} \cdot \frac{dk_{SH}}{dz}, \qquad (4)$$

where $dz = Rd\theta$, ($R$ is the radius of the microresonator), and $\Delta k = 2k_F - k_{SH}$ represents the difference between the wave vector of the fundamental wave $k_F = 2\pi n_F/\lambda_F$ and that of the second harmonic wave $k_{SH} = 2\pi n_{SH}/\lambda_{SH}$. In addition, $\omega_{SH}$, c, $n_{SH}$, and $E_F$ are the circular frequency of the second harmonic, light speed in vacuum, $\theta$-dependent effective refractive index of the second harmonic, and field strength of the fundamental wave circulated in the microresonator, respectively. The gain dynamics of SHG was investigated by integrating Eq. (4) along the periphery over a single cycle in the steady-state regime. The details on the derivation of

Eq. (4) can be found in **Supplementary Material**.

## Acknowledgements

The work is supported by National Basic Research Program of China (No. 2014CB921300), NSFC (Nos. 61275205, 11174305, 61405220, and 61505231), and the Fundamental Research Funds for the Central Universities.

## Author contributions

J. L., W. F. and Y. C. planned and designed the experiments. J. L., Z. F. and M. W. established the experimental setup. J. L. and M. W. fabricated the microresonator using femtosecond laser direct writing. J. L. and Y. X. performed FIB milling. J. L., L. Q., M. W. carried out the SHG experiment. J. N. and J. L. performed the numerical simulation of the microresonator. J. L., W. F. and Y. C. analysed data and wrote the manuscript.

**Figure Captions:**

Fig. 1 Experimental setup for the SHG in the LN microresonator. Polarization states of TE and TM were schematically illustrated near the microresonator. The pump and second harmonic signals coupled out of the microresonator with the fiber taper were collimated by an objective lens with an NA of 0.25. The inset on the left-hand side: experimental setup of determining polarization states of the light beams scattering from the edge of the microresonator, where $\theta$ is the angle between the wave vector and optical axis of the crystal; the inset on the right-hand side: SEM image of the microresontor. Also we indicate the X, Y, and Z directions in the inset on the left-hand side. It should be noted that the setup illustrated in the inset on the left-hand side is different from that illustrated in the main figure for SHG.

Fig. 2(a) The spectrum of the TE-polarized second harmonic generated from the microresonator. (b) The spectrum of the TM-polarized pump light. (c) Side-view optical micrograph of the second harmonic scattering from the edge of microresonator, where the red rectangle depicts the contour of the microdisk. (d) The SHG conversion efficiency plotted as a function of the pump power.

Fig. 3 (a) Transmission spectrum around 1540 nm recorded by coupling the fiber taper with the microresonator. Inset: Lorentzian fit (red solid line) of measured spectrum around the resonant wavelength at 1540.13 nm (black dotted line), showing a Q factor of $1.13\times10^5$. (b) Transmission spectrum around 770 nm recorded by coupling the fiber taper with the microresonator. Inset: Lorentzian fit (red solid line) of measured spectrum around the resonant wavelength at 770.065 nm (black dotted line), showing a Q factor of $1.6\times10^4$.

Fig. 4 Numerically calculated (a) effective refractive indices of the pump (solid blue curve) and second harmonic modes (red dashed curve), (b) difference in the $k$ vectors of the pump and second harmonics modes ($\varDelta k$), (c) nonlinear coefficient, and (d) electric field amplitude of second harmonic signal versus the azimuth angle along the microresonator periphery.

Fig. 1

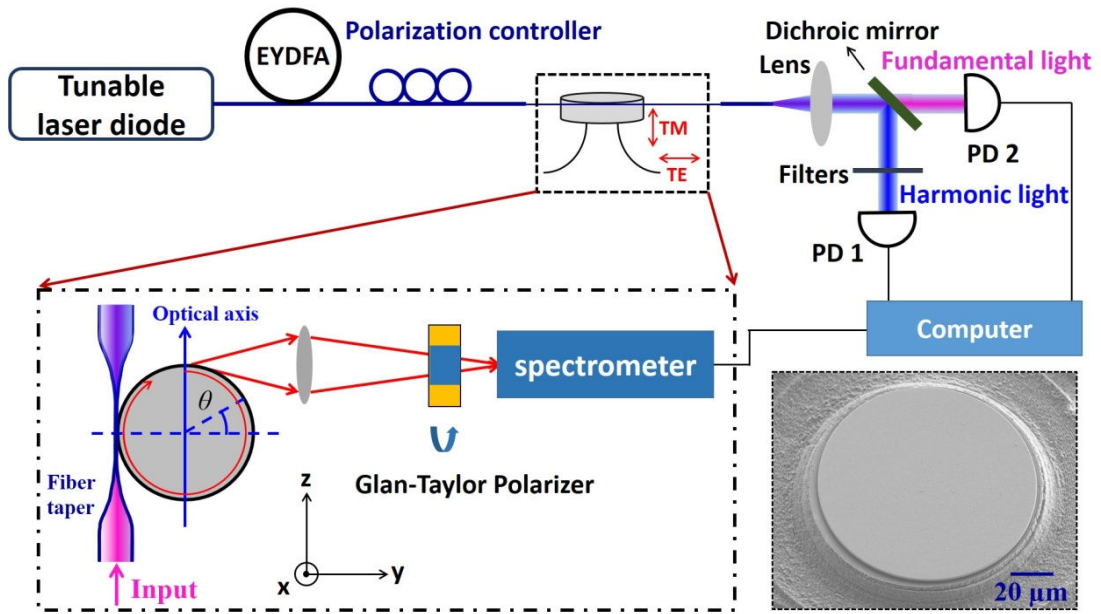

Fig. 2

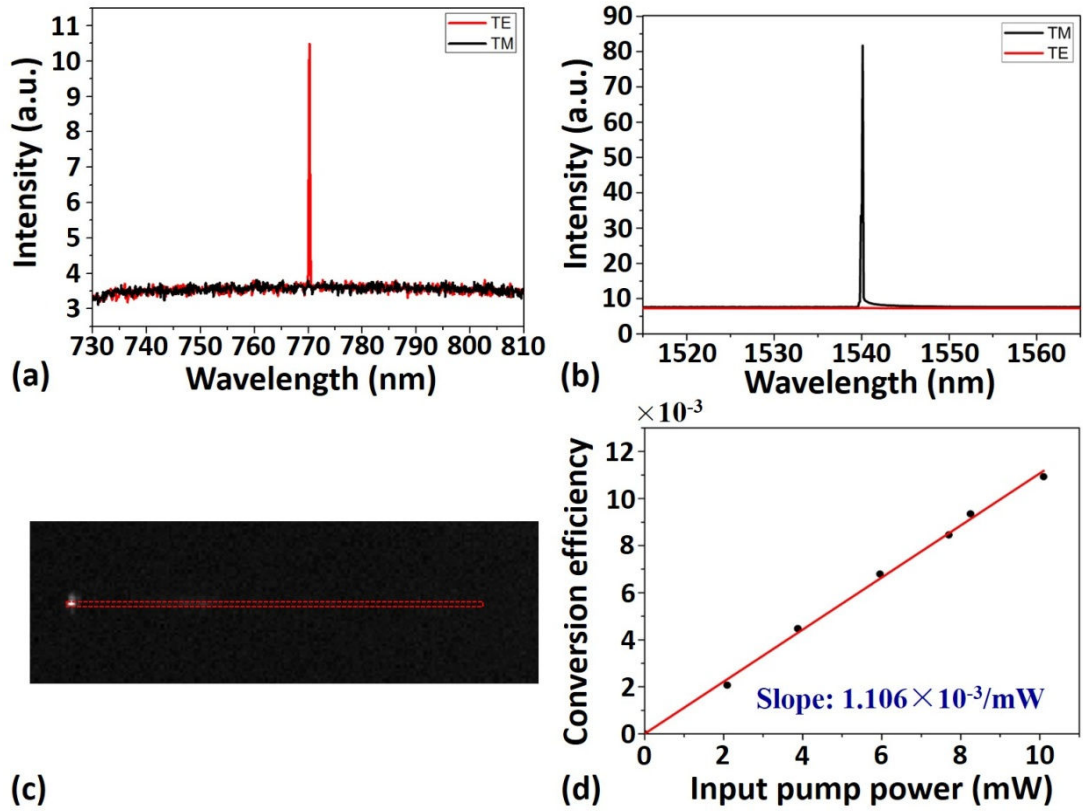

Fig. 3

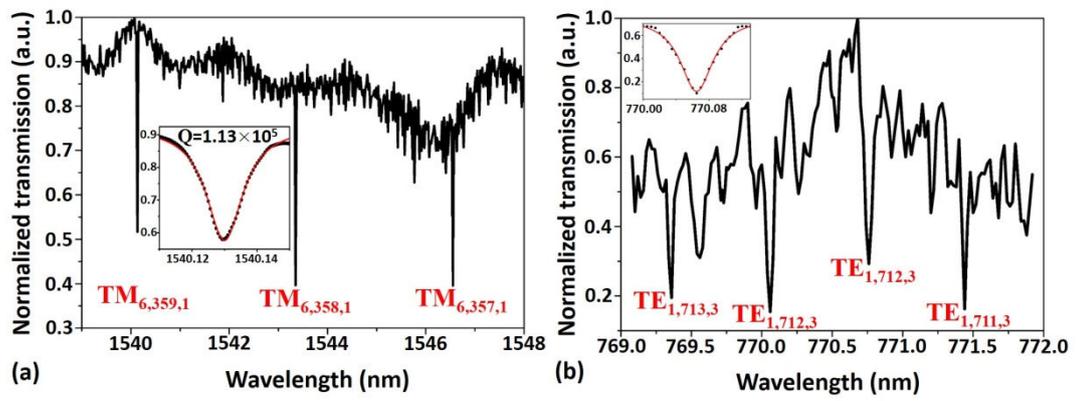

Fig. 4

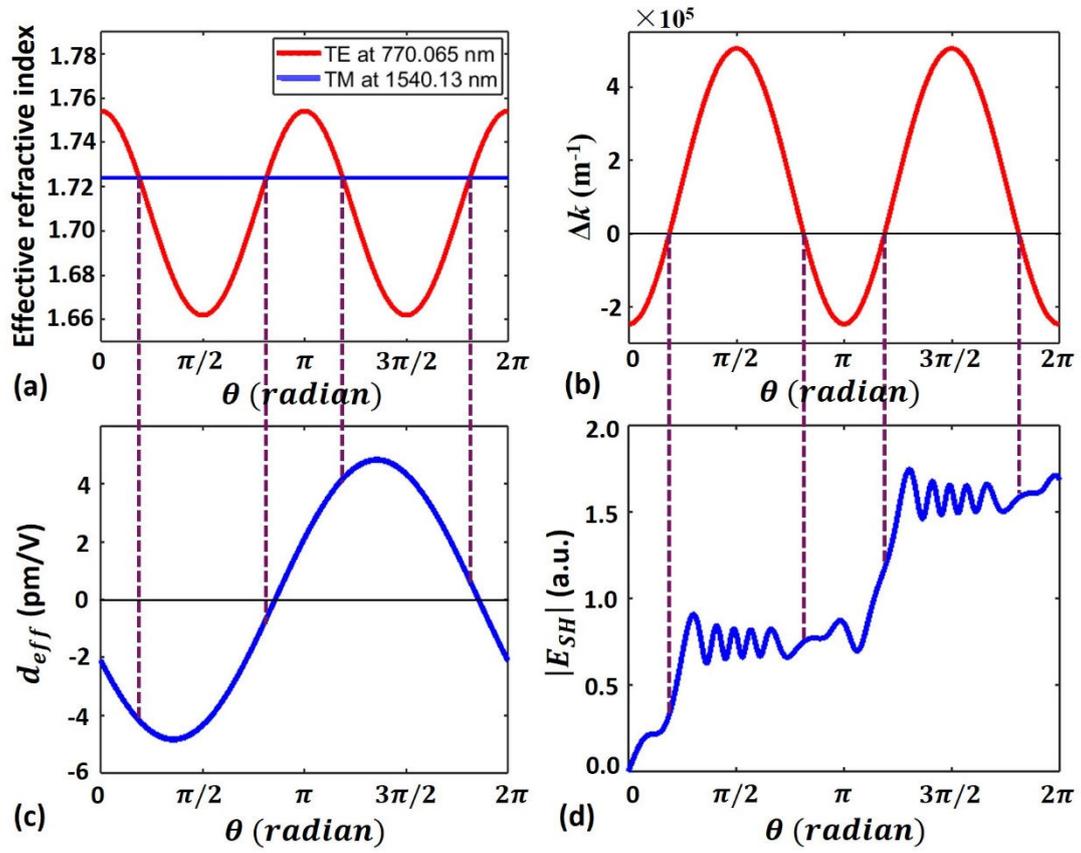

# Supplementary material

To derive Eq. (4) in the manuscript, we begin with the coupled-wave equations for sum-frequency generation in Chapter 2.2 of *Nonlinear Optics* authored by Robert W. Boyd (Third edition) [S1].

Since we are discussing second harmonic generation, a plane wave at the second harmonic frequency $\omega_{SH}$ propagating in the +z direction is

$$\tilde{E}_{SH}(z,t) = E_{SH}(z)e^{i\int k_{SH}(z)dz - i\omega_{SH}t} + c.c., \qquad (S1)$$

where

$$k_{SH}(z) = \frac{n_{SH}(z)\omega_{SH}}{c}, n_{SH}^2 = \epsilon^{(1)}(\omega_{SH}),$$

and $E_{SH}$ is the amplitude of the second harmonic.

Based on Maxwell equations, we have

$$\nabla^2 \tilde{E}_{SH} - \frac{\epsilon^{(1)}(\omega_{SH})}{c^2}\frac{\partial^2 \tilde{E}_{SH}}{\partial t^2} = \frac{1}{\epsilon_0 c^2}\frac{\partial^2 \tilde{P}_{SH}}{\partial t^2}. \qquad (S2)$$

$\tilde{P}_{SH}$ is the second order polarization, which appears as

$$\tilde{P}_{SH}(z,t) = P_{SH}e^{-i\omega_{SH}t} + c.c., \qquad (S3)$$

where

$$P_{SH} = 4\epsilon_0 d_{eff}\tilde{E}_F^2(z), \ \tilde{E}_F(z) = E_F e^{ik_F z} + c.c.$$

Note that the overall conversion efficiency of our second harmonic generation experiment is quite low (i.e., ~10$^{-3}$), therefore the depletion of pump beam is neglected. In addition, with the plane wave assumption, the second harmonic field depends only on the coordinate z. Then we have

$$\nabla \tilde{E}_{SH} = \frac{dE_{SH}}{dz} \cdot e^{i\int k_{SH}(z)dz - i\omega_{SH}t} + ik_{SH}(z)E_{SH}(z) \cdot e^{i\int k_{SH}(z)dz - i\omega_{SH}t} + c.c.,$$

$$\nabla^2 \tilde{E}_{SH} = \left[\frac{d^2 E_{SH}}{dz^2} + 2ik_{SH} \cdot \frac{dE_{SH}}{dz} + iE_{SH} \cdot \frac{dk_{SH}}{dz} - k_{SH}^2 E_{SH}\right] \cdot e^{i\int k_{SH}dz - i\omega_{SH}t} + c.c.. \qquad (S4)$$

We substitute Eqs. (S1), (S3), and (S4) into the wave equation (S2), and then obtain

$$\left[\frac{d^2E_{SH}}{dz^2} + 2ik_{SH}\frac{dE_{SH}}{dz} + iE_{SH}\frac{dk_{SH}}{dz} - k_{SH}^2 E_{SH} + \frac{\epsilon^{(1)}(\omega_{SH})\omega_{SH}^2}{c^2}E_{SH}\right]e^{i\int k_{SH}dz - i\omega_{SH}t} + c.c.$$

$$= \frac{-4d_{eff}\omega_{SH}^2}{c^2}E_F^2 e^{i(2k_F z - \omega_{SH}t)} + c.c. \tag{S5}$$

Since $k_{SH}^2 = \frac{\epsilon^{(1)}(\omega_{SH})\omega_{SH}^2}{c^2}$, the fourth and fifth terms in bracket of Eq. (S5) cancel out. The above Eq. (S5) can be rewritten as

$$\frac{d^2E_{SH}}{dz^2} + 2ik_{SH} \cdot \frac{dE_{SH}}{dz} + iE_{SH} \cdot \frac{dk_{SH}}{dz} = \frac{-4d_{eff}\omega_{SH}^2}{c^2}E_F^2 e^{i(2k_F z - \int k_{SH}dz)}.$$

For the X-cut crystal microdisk, the pump wave has a constant $k_F$. We set $\Delta k = 2k_F - k_{SH}$, and we obtain

$$\frac{d^2E_{SH}}{dz^2} + 2ik_{SH} \cdot \frac{dE_{SH}}{dz} + iE_{SH} \cdot \frac{dk_{SH}}{dz} = \frac{-4d_{eff}\omega_{SH}^2}{c^2}E_F^2 e^{i\int \Delta k \cdot dz}. \tag{S6}$$

Under the slowly varying amplitude approximation, namely $\left|\frac{d^2E_{SH}}{dz^2}\right| \ll \left|k_{SH}\frac{dE_{SH}}{dz}\right|$, Eq. (S6) becomes

$$\frac{dE_{SH}}{dz} = \frac{2id_{eff}\omega_{SH}^2}{k_{SH}c^2}E_F^2 e^{i\int \Delta k \cdot dz} - \frac{E_{SH}}{2k_{SH}} \cdot \frac{dk_{SH}}{dz}. \tag{S7}$$